# The Mn-O bonds dependence of the lattice distortion in LaMnO$_3$

**Mahrous r Ahmed[1], A. M. Ahmed[1], A. K. Diab[1], S. M. Abo-elhasan[1] and G A Gehring[2]**

[1]*Department of Physics, Faculty of Science, University of Sohag, Sohag 82534, Egypt.*

[2]*Department of Physics and Astronomy, University of Sheffield, Hicks Building, Hounsfield Road, Sheffield S3 7RH, United Kingdom.*

## Abstract

We have investigated the volume collapse occurring in LaMnO$_3$ unit cell using the anisotropic Potts model modified by two types of anisotropic interactions which has been used to study the change of Mn-O bonds lengths as function of temperature. The bond lengths are related to the occupation of the orbits. The lattice parameters and the distortion modes have been investigated as well. We have shown that the collapse is due to the change of Mn-O bonds lengths change as the temperature is raised through the transition. It has been shown that all the parameters studied here decrease with increasing temperature in a narrow temperature range below $T_{JT}$, and then undergoes a collapse at $T_{JT}$. These results are in a good agreement with the published experimental results.

# I.Introduction:

Hole-doped LaMnO$_3$, La$_{1-x}$A$_x$MnO$_3$ *(A*=Ca, Sr, and Ba), exhibits colossal magnetoresistance and has attracted very much interest through the last 50 years.[1,2] For this reason a large number of theoretical investigations has already been undertaken on stoichiometric LaMnO$_3$ itself.[3].

In such materials each unit cell contains one (or more) localized magnetic moments usually called spins. At low temperature interactions between spins cause long range order. This long range order disappears at a critical temperature, T$_C$. Various theoretical approaches are used to study such materials. Simulations are used for Ising model and classical spin models, simulations are harder to use for quantum spins.

Recently much interest has been attracted to understand a new form of the order, namely, the orbital ordering which has been seen in $V_2O_3$, $LaVO_3$ and also in $LaMnO_3$. Much work have been done on understanding both the nature and the mechanism of orbital ordering in $LaMnO_3$ but very little work has been done on the temperature dependence of such ordering[4,5,6].

Orbital ordering is always accompanied by lattice distortions. Jahn-Teller (JT) effect lifts the degeneracy of the 3d doublet orbitals, $e_g$, by a cooperative JT lattice deformation. This involves a compression of the Mn-O octahedra along c-axis, while alternate JT deformations occur in the ab-plane which stabilise a certain type of orbital ordering. $LaMnO_3$ crystallizes in the orthorhombic space group $Pb_{nm}$ and undergoes a transition at $T_{JT}$ = 750 K from that Jahn-Teller distorted orthorhombic phase to a high-temperature orthorhombic phase which is nearly cubic[7]. The transition is accompanied by an orbital order-disorder transition. The low-temperature phase has three Mn-O bond lengths called short (*s*), medium (*m*), and long (*l*).

Chatterji *et al*[8] discovered an unusual abrupt volume contraction at $T_{JT}$ =750 K. The high-temperature phase just above $T_{JT}$ has less volume than the low-temperature phase. This volume contraction is very rare in solid-solid structural phase transition. A well known example is the volume contraction in Fe[10] at the bcc-to-fcc (α-Fe to γ–Fe) structural transition at $T_C$ = 1185 K.

Chatterji *et al.*[8] investigated the Jahn-Teller transition in $LaMnO_3$ using high-temperature x-ray and neutron diffraction on powder samples. They observed that the unit-cell volume of $LaMnO_3$ decreases with increasing temperature in a narrow temperature range below $T_{JT}$ and then undergoes a sudden collapse at $T_{JT}$. It was argued that this striking volume collapse is caused by the orbital order-disorder transition. In the orbitally ordered phase the packing of $MnO_6$ octahedra needs more space than in the disordered phase.

Bozin et al.[9] also obtained the volume contraction for the same compound as function of doping range $x$=0.0 to 0.5 at fixed $T$=550 K and as function of temperature at fixed doping $x$=0.0. They investigated their compound by advent of high data throughput neutron powder diffractometers.

Theoretically, Maitra et al.[11] constructed a model Hamiltonian involving the pseudospin of $Mn^{3+}$ $eg$ states, the staggered JT distortion and the volume strain coordinate to study the volume collapse of $LaMnO_3$ at the JT transition temperature. They also have shown that the anharmonic coupling between these primary- and secondary-order parameters leads to the first-order JT phase transition associated with a comparatively large reduction of the unit-cell volume.

Millis[12] derived a classical model, which was based on previous work by Kanamori[13] for the lattice distortions in manganites. The model may be approximated either by an antiferromagnetic $xy$ model with a modest threefold anisotropy or by a three-state Potts model with an antiferromagnetic first-neighbor interaction and a weak second-neighbor interaction. This differs from our model which deals only with the nearest-neighbor interaction.

Ahmed and Gehring[6] have investigated the volume collapse occurring in $LaMnO_3$ unit cell using the anisotropic Potts model modified by two types of anisotropic interactions which was used to study the behavior of the Mn-Mn orbital ordering configurations with temperature. The theoretical volume obtained by calculating of Mn-Mn orbitals ordering probability was found equal to 1/4 of the experimental volume where the lattice distortion was considered in the experimental results and a unit cell four times bigger than the unit cell was used. That model needed more modifications to be able to calculate the behavior of $LaMnO_3$ lattice distortion parameters and obtain a correct volume with the temperature.

The Mn-O bond lengths are related to the occupation of the orbits. In this paper we modify the anisotropic Potts model used in [6] to count the $l$, $m$ and $s$ Mn-O bonds probability to investigate the volume collapse occurring in $LaMnO_3$ unit cell as function of temperature. We have shown that the collapse of the volume due to the change of Mn-O bonds lengths as the temperature is raised through the transition has a first order transition.

The two distortion modes are expressed in terms of *l*, *s*, and *m*. The Mn-O bonds and two octahedral distortion modes, $Q_2$ and $Q_3$ have been studied as function of temperature. The results obtained from this work for the temperature dependence of bond lengths, lattice parameters, unit cell volume and the distortion modes $Q_2$ and $Q_3$ for the $LaMnO_3$ agree with what were published experimentally[8].

We explain the model in brief in Sec. II and the Monte Carlo, MC, simulation condition in Sec. III. The results are discussed in Sec. IV and finally are concluded in Sec. V.

## II. The model

A three-state Potts model in three dimensions allows for the possibility that the Potts states and the space coordinates are coupled. This arises physically where there is strong JT coupling of an electronic doublet typically from *d* electrons coupled to the two-dimensional lattice distortions with strong unharmonic terms as discussed by Kanamori[14].

Because the orbital ordering in the transition metal oxides has three anisotropic states it is natural to set up a three states Potts model. The standard q-state Potts model[14] consists of a lattice of spins, which can take *q* different values from 1 to *q*, here q=3, and whose Hamiltonian is

$$H = -\frac{J}{2} \sum_{<i,j>}^{N} \delta_{S_i, S_j} \qquad (1)$$

where $S_i$=1,2, . . . is one of the *q* states on site *i*, $\delta_{S_i,S_j}$ is the Kronecker function which is equal to 1 when the states on sites *i* and *j* are identical, $S_i=S_j$, and is zero otherwise, <i , j> means that the sum is over the nearest-neighbor pairs, J is the exchange integral and *N* is the total number of sites in the lattice. For *q*=2, this is equivalent to the Ising model. The Potts model is, thus, a simple extension of the Ising model, however, it has a much richer phase structure, which makes it an important testing ground for new theories and algorithms in the study of critical phenomena.[15]

The anisotropic three-state Potts model on a simple cubic lattice is given as

$$H_{AIS} = -\frac{1}{2}\sum_{<ij>} J_{ij}(\rho_{ij})\delta_{ij}$$  (20

where $\rho_{ij}=R_i-R_j$ is the bond distance between $i$ and $j$ sites. The factor of 1/2 is included to correct for double counting. In such a model the interaction between orbits depends on both the type of the orbits and the bond direction between each two of them.[16] The compound $LaMnO_3$ has orbital ordering of this type, as seen in Fig. 1, and the interactions between the orbits have been calculated.[17,18,19]

If we start our simulation from high temperature we obtain any of the three possible configurations for the orbital-ordering ground state. The symmetry is broken at $T_C$ so that the Monte Carlo simulations give one of the possible ground states and not an average of all of them.

## III. Monte Carlo simulation

This model has been studied using MC simulations on 3d finite lattices with periodic boundary conditions (with size $L^3$ where $L$=12). All our simulations have made use of the Metropolis algorithm with averaging performed over from $10^5$ to $10^7$ Monte Carlo steps per site. Results were obtained by either cooling down from a high-temperature random configuration as discussed by Banavar *et al.*[20] or heating up from the ground state. The results from the two procedures agree.

## IV. Results and Discussion

In a previous work[2] we used the modified Potts model to obtain the orbital ordering in $LaMnO_3$ and in another paper[3] we studied the volume collapse occurring due to the Mn-Mn orbital ordering configurations probability in $LaMnO_3$ unit cell. The occupation of the orbitals on Mn neighboring sites affects the Mn-Mn separation and hence Mn-O bond, as shown in Fig. 2, which either is long, *l*, medium, *m*, or short, *s*, bond. We obtained the occupations of the orbits from the Monte Carlo simulation and hence the probabilities for the Mn-O bond as shown in Fig. 2. The orbital ordering in the ground state of $LaMnO_3$ is shown in fig. 1. We now use the bond lengths to relate these to the volume. The experimental[7] values of $l$, $m$, and $s$ are as the following: $l$=2.178 Å, $m$ = 1.968 Å, and $s$ = 1.907 Å.

We note that the lengths of the short and medium bonds differ slightly whereas the long bond is significantly longer. The lobe of the Mn d-orbital is pointing away from the oxygen in the short bond of the orthogonal, the side-side and parallel-orthogonal configurations[13]. We assume that this is because of the Coulomb repulsion between the d-orbitals in the side-side configuration and so choose the bond length for the parallel orthogonal to equal that of the short bond, s.

Now we calculate the probability of occurring Mn-O bonds between the two Mn-Mn orbitals. The two bonds which are between the two orbitals of the orthogonal configuration are $l$ and $s$, see fig. 2a, between the two orbitals of the side-side configuration are $m$ and $m$, see fig. 2b, and between the orbitals of the parallel-orthogonal configuration are two shorts, $s$ and $s$, see fig. 2c. With the temperature raising another configuration appears with two long bonds, $l$ and $l$, called head-head as shown in fig. 2d.

In each $MnO_6$ octahedra, Mn-site has six oxygen, O, nearest neighbors. We calculate the probabilities of occurring of the long, short and medium bonds along the six directions ±x-, ±y- and ±z-axes. The average for every Mn-O bond along the six directions is calculated. The average of the long bond probability, <long>, for example, along the six axes of the $MnO_6$ octahedra for the whole lattice is as following,

<long>=[<long>$_{px}$+<long>$_{nx}$+<long>$_{py}$+<long>$_{ny}$+<long>$_{pz}$+<long>$_{nz}$]/n    (3)

Where $p_x$ and $n_x$, $p_y$ and $n_y$ and $p_z$ and $n_z$ are the positive and negative direction of the x-, y- and z-axis respectively. Similarly we are able to obtain the average of the medium probability, <medium>, and short probability, <short>, bonds. Note that

n = 6. Because at T = 0 the long bonds probability equals 1/3 of the whole probabilities, see fig. 1, we multiply equation (3) in 3.

$$l = 3*<long>*2.178, \qquad (4)$$
$$\text{similarly } m = 3*<medium>*1.968 \qquad (5)$$
$$\text{and } s = 3*<short>*1.907 \qquad (6)$$

Figure 3. shows the temperature dependence of the short $s$, long $l$, and medium $m$. The long and short Mn-O bond probabilities have a first order transition at the Jahn-Teller temperature, $T_{JT}$, where the medium Mn-O bonds probability nearly does not change with the temperature. After the transition temperature the $l$, $m$ and $s$ are nearly equal and unchangeable with the temperature because the crystal is pseudo cubic at high temperature.

The temperature variation of the lattice parameters $a$, $b$, and $c/\sqrt{2}$ is shown in fig. 4. The lattice parameter $b$ decreases, the parameter $a$ nearly does not change along the temperature range while the parameter $c/\sqrt{2}$ increases with temperature. All the parameters abruptly become almost the same (metrically almost cubic) at $T_{JT}$. The transition is clearly first order, because in a short temperature range both high and low temperature phases coexist.

We use $l$, $m$ and $s$ obtained above to calculate the volume, $V(T)$, of LaMnO$_3$ unit cell in the following equation.

$$V(T) = (l + s)(l + s)(m + m) = 2m(l + s)^2, \qquad (7)$$
$$\Delta V = |V(0) - V(T)| \qquad (8)$$

Figure 5 shows the temperature variation of the absolute value of $\Delta V$ which is obtained after subtracting $V(T)$, obtained by equation (7), from the cell volume base $V(T=0)$. The temperature variation of this extra volume $\Delta V$ looks like an order parameter that decreases continuously at first and then drops abruptly to zero at $T_{JT}$. These results agree very well with what was obtained experimentally by Chatterji *et al*[8] with an error nearly equal $\approx 0.16$ at T = 0. This result is considered a modification to our theoretical results published in ref. [6] for $\Delta V$.

Sa´nchez et al.[21] measured the heat capacity anomaly of LaMnO$_3$ at T$_{JT}$. The enthalpy involved in this transition is found to be ΔH=3180±100 J mole$^{-1}$. This value of ΔH combined with the volume drop at T$_{JT}$, ΔV=-0.1535 cm$^3$ mole$^{-1}$, determined during the present investigation yields from the Clapeyron equation ΔT/ΔP = -3.62 Kkbar$^{-1}$. S´nchez *et al.* interpreted the transition at T$_{JT}$ as an order-disorder transition where the statistical occupation in the disordered phase corresponds to the three possible orientations of the tetragonal distortions on each octahedron. They claimed that the three-state Potts model is the appropriate statistical model for this transition. However, the calculated value of entropy[5] by our anisotropic Potts model at T$_{JT}$ is more accurate than that was obtained experimentally by their model.

Because the distortion of the MnO$_6$ octahedron result in the Jahn-Teller effect produces three Mn-O bond distances: long (*l*), short (*s*), and medium (*m*) the distorted crystal structure can be obtained from the ideal perovskite structure by the following way: first the distortion Q$_2$ of the octahedron formed with O$^{-2}$ ions is added in a staggered way along the three directions, and then the distortion Q$_3$ is superimposed on it. These two distortion modes are expressed in terms of *l, s,* and *m* by

$$Q_2 = \frac{2}{\sqrt{2}}(l-s) \quad (9)$$

$$Q_3 = \frac{2}{\sqrt{6}}(2m-l-s) \quad (10)$$

Figure 6 shows the temperature dependence of the two distortion modes Q$_2$ and Q$_3$. Q$_2$ starts with higher value than that for Q$_3$ value because it almost depends on the long bond where Q$_2$ has very small value at T = 0. Both Q$_2$ and Q$_3$ also have a first order transition at T$_{JT}$ and have the same value above the transition temperature because of disappearing of the lattice distortion. However, the thermal parameters of the Oxygen-atoms become very large in the high temperature orthorhombic and the rhombohedral phases.

# V. Conclusions

The Potts model with two anisotropic orbital interaction types has produced the orbital-ordering phase which occurs in $LaMnO_3$ at specific values of the interaction types. The internal energy and specific heat for the same phase were obtained also from the same model.[4] The model also has produced good results for the order parameter and for the entropy at the orbital-ordering phase-transition temperature $T_{JT}$ confirmed by experiment in ref.[22] and ref[21] respectively.

In this paper, we used the same model with more modifications to include the relation of the orbital ordering-disordering to the bond lengths. The model modifications made us able to calculate the behavior of the temperature variation of the Mn-O bonds distances, $l$, $m$ and $s$. Also, we calculated the lattice parameters, volume and the two distortion modes $Q_2$ and $Q_3$ change with the temperature that occurs at $T_{JT}$ in $LaMnO_3$ unit cell which have been studied by Chatterji *et al.*[9] It was found that all these physical quantities calculated for the $LaMnO_3$ unit cell have a first-order transition at $T_{JT}$.

This model showed that this is due to the remarkable decrease in number of the long Mn-O bonds and a less increase in the number of the short Mn-O bonds as the lattice is warmed through $T_{JT}$. The theoretical volume which is calculated here was found in agreement with the experimental volume with an error equals to $\approx 0.16$ at T = 0 where the lattice distortion is involved here as well as was considered in the experimental results.

In the future work, we will try to do more modifications on our anisotropic Potts model to be able to apply it widely on other metal-insulators transition compounds.

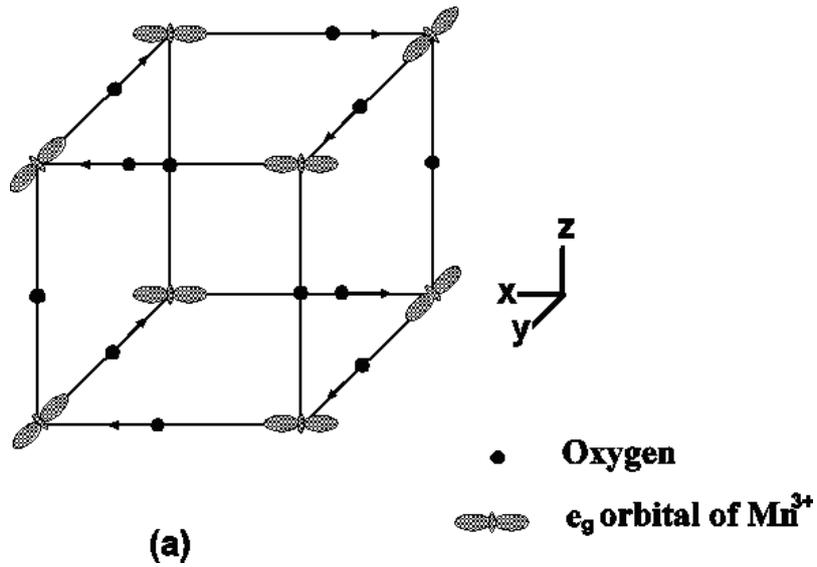

Fig. (1): The orbital ordering in LaMnO$_3$. It is antiferromagnetic in the *x-y* plane and ferromagnetic along the *z*-axis. Each Mn-site has six Oxygen, O, nearest neighbours with two long Mn-O bonds, two medium Mn-O bonds and two short Mn-O bonds at T= 0.

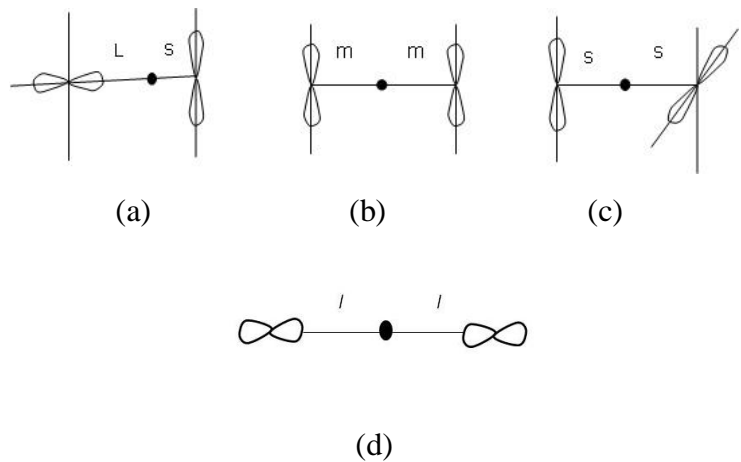

Fig. (2): The three different Mn-O bonds *l*, *m*, and *s* are shown in the four different Mn-O-Mn orbital ordering configurations. Note that the orbits are Mn-sites and the black solid circles are the Oxygen-sites.

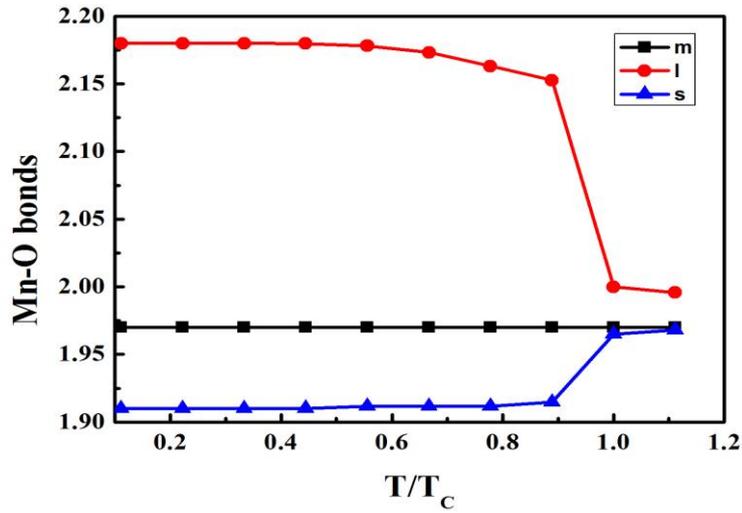

Fig. (3): Temperature variation of the three Mn-O bond lengths *l*, *m* and *s* of LaMnO$_3$.

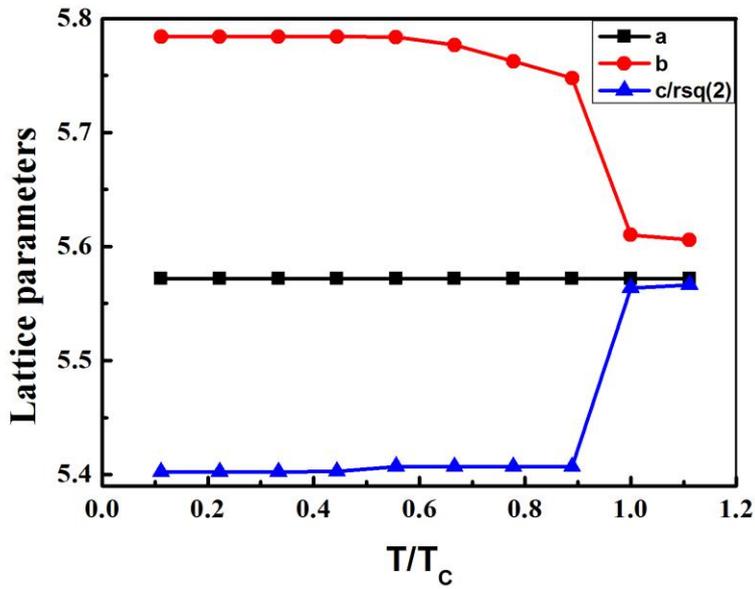

Fig. (4): Temperature variation of the lattice parameters of LaMnO$_3$. The error bars are smaller than the sizes of the data symbols.

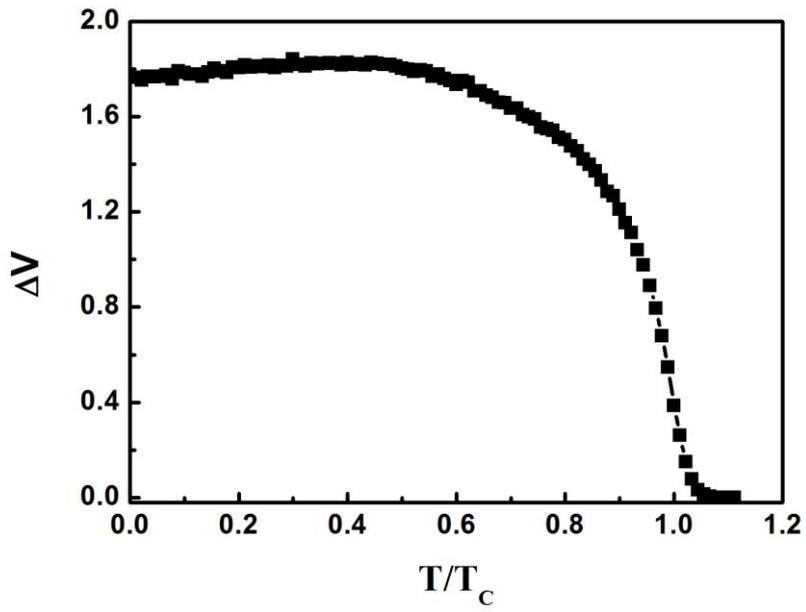

Fig. (5): Temperature variation of the unit cell volume after subtraction of the base volume explained in the text.

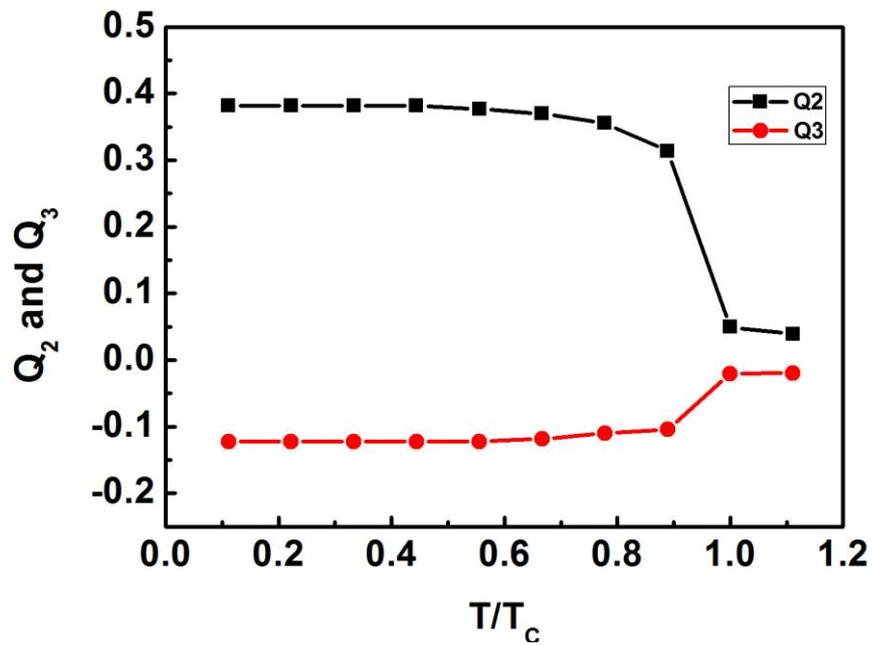

Fig. (6): Temperature variation of the two octahedral distortion modes, $Q_2$ and $Q_3$ defined in Eq. ( ).